\documentstyle[pra,preprint,aps]{revtex}

\begin{document}
\draft
\title{Minimum-error discrimination between symmetric mixed quantum states}
\author{C.-L. Chou\thanks{%
choucl@cycu.edu.tw} and L. Y. Hsu \thanks{%
lyhsu@phys.cts.nthu.edu.tw}}
\address{$^*$Department of Physics, Chung-Yuan Christian University,\\
Taoyuan, Taiwan 32023}
\address{$^\dag$Physics Division, National Center of Theoretical Science,\\
Hsinchu, Taiwan 30055}
\date{\today}
\maketitle

\begin{abstract}
We provide a solution of finding optimal measurement strategy for
distinguishing between symmetric mixed quantum states. It is
assumed that the matrix elements of at least one of the symmetric
quantum states are all real and nonnegative in the basis of the
eigenstates of the symmetry operator.
\end{abstract}




\section{Introduction}
The theory of quantum information and communication is a
well-developed field of research \cite{Qdetect,Holevo,Qinfo}. It
concerns the transmission of information using quantum states and
channels. The transmission party encodes a message onto a set of
quantum states $\{\rho_k\}$ with prior probability $p_k$ for each
of the states $\rho_k$. The set of signal states and the prior
probabilities are also known to the receiving party. The task of
the receiving party is to decode the received message, i.e.,
finding the best measurement strategy based upon the knowledge of
the signal states and their prior probabilities. One possibility
is to choose the strategy that minimizes the probability of
detection error. In this paper we will consider the minimization
of the probability of error for a certain class of quantum
ensembles.

In general, the measurement strategy is described in terms of a
set of nonnegative-definite operators called the probability
operator measure (POM) \cite{Qdetect,Holevo}. The measurement
outcome labeled by $"k"$ is associated with the element ($\pi_k$)
of POM that has all the eigenvalues be either positive or zero.
The POM elements must sum into the identity operator $\sum_k \pi_k
= \hat{\mathbf{1}}$. The probability that the receiver will
observe the outcome $k$ given that the transmitted signal is
$\rho_j$ is $P(k|j)={\verb"tr"}(\pi_k\rho_j)$. Here \verb"tr"
denotes the trace operation. It follows that the error probability
is given by
\begin{equation}
P_{error}=1-\sum_{k}p_k\verb"tr"(\pi_k\rho_k).
\label{eqn:errorProb}
\end{equation}

The necessary and sufficient conditions that lead to the minimum
error probability are known to be
\cite{Qdetect,Holevo,Yuen,Symmetric}
\begin{eqnarray}
\pi_k(p_k\rho_k - p_j\rho_j)\pi_j&=&0, \label{eqn:opRelation}\\
\sum_{k}p_k \pi_k\rho_k -p_j\rho_j &\geq& 0 .
\label{eqn:positivity}
\end{eqnarray}
\noindent The first condition holds for all $j$ and $k$. The
second condition means that all the eigenvalues of the operator at
the left-hand side are nonnegative and it holds for all $j$. These
conditions are highly nontrivial such that the required POM
elements for the best measurement strategy are not easily derived
from the conditions. In fact, only some classes of quantum
ensembles are known for their best measurement strategies. These
include the cases of only two signal states \cite{Qdetect},
symmetric states \cite{Symmetric,Multiply}, mirror-symmetric
states \cite{mirror}, linearly independent states \cite{linear},
and equiprobable states that are complete in the sense that a
weighted sum of projectors onto the states equals the identity
operator \cite{Yuen}.

In this paper, we will provide the optimal measurement strategy
for a set of $N$ mixed symmetric quantum states $\{\rho_k\}$.
These states are of equal prior probabilities $p_k= 1/N$ and
assumed to respect the $Z_N$ symmetry
\begin{eqnarray}
\rho_k &=& R^k\rho_0R^{\dagger k},\hspace{1cm} k=0,1,...,(N-1),
\label{eqn:rho}\\
R^N &=& \pm \hat{\mathbf{1}},\label{eqn:RN1}
\end{eqnarray}
\noindent where the operator $R$ denotes the relevant part of the
symmetry operator that lives in the same Hilbert subspace of the
signal states $\{\rho_k \}$. $\hat {\mathbf{1}}$ denotes the
identity operator of the Hilbert subspace of the signal states. We
also assume $R$ to be unitary $(R R^\dagger=R^\dagger
R=\hat{\mathbf{1}})$ and nondegenerate, i.e., all its eigenvalues
$\{b_\lambda\}$ are different for different eigenstates
$\{|\lambda\rangle\}$. Therefore the dimensionality of $R$ cannot
be larger than the number of the signal states $N$ otherwise at
least two of the eigenvalues of $R$ will be the same. Besides, we
also assume that at least one of the the signal states (assigned
to be $\rho_0$) can be made to have all it matrix elements be real
and nonnegative, i.e., $\langle \lambda | \rho_0 | \lambda'
\rangle \geq 0$ for some chosen set of the eigenstates $\{|\lambda
\rangle \}$ of the operator $R$.

\section{The Optimal Measurement Strategy}
In many of the cases \cite{Symmetric,Smixed} where the optimal
strategies are known to be the square-root measurements with POM
elements
\begin{eqnarray}
\pi_k &=& \Phi^{- {1\over 2}}(p_k \rho_k) \Phi^{-{1\over
2}},\label{eqn:squareroot}\\
\Phi&\equiv&\sum_{k}p_k\rho_k
\end{eqnarray}
\noindent where $\rho_k$ denotes the $k$-th quantum signal states
to be discriminated, and $\Phi$ is invariant under the
transformation $R$. In this paper, we assume the invariant
operator $\Phi$ as
\begin{equation}
\Phi\equiv \sum_{k=0}^{N-1} R^k \Gamma_0 R^{\dagger k},
\label{eqn:definePhi}
\end{equation}
\noindent where $\Gamma_0\equiv |\varphi_0 \rangle
\langle\varphi_0|$ is the rank one operator that is formed by some
normalized pure quantum state $|\varphi_0\rangle$. From equation
(\ref{eqn:definePhi}), $\Phi$ is Hermitian and
nonnegative-definite, and commutes with $R$. This implies that
both $R$ and $\Phi$ can be expanded in terms of the same
orthonormal basis $\{|\lambda\rangle \}$ as
\begin{eqnarray}
\Phi&=&\sum_{\lambda} a_{\lambda}|\lambda\rangle\langle \lambda|,
\label{eqn:Phiinlambda}\\
R&=&\sum_{\lambda}b_{\lambda}|\lambda\rangle\langle \lambda|,
\label{eqn:Rinlambda}
\end{eqnarray}
\noindent where $a_{\lambda} = N|\langle \lambda | \varphi_0
\rangle|^2 $ for all $\lambda$. In general, it is difficult to
obtain the POM elements that satisfy the conditions
(\ref{eqn:opRelation}) and (\ref{eqn:positivity}). However, we can
obtain a solution to these conditions for the symmetric mixed
quantum states described in the equations (\ref{eqn:rho}) and
(\ref{eqn:RN1}).

{\it Proposition}. Given the mixed symmetric quantum states as
described in the equations (\ref{eqn:rho}) and (\ref{eqn:RN1}),
the optimal measurement strategy that minimizes the error
probability $P_{error}$ is described by the POM $\{\pi_k\}$ that
is defined by
\begin{eqnarray}
\pi_k &\equiv& R^k \Phi_2 \Gamma_0 \Phi_2 R^{\dagger k},
\hspace{1cm}
k=0,1,..,N-1. \label{eqn:POM}\\
\Gamma_0 &\equiv& |\varphi_0 \rangle \langle \varphi_0|,
\label{eqn:Gamma}
\end{eqnarray}
\noindent where $|\varphi_0\rangle$ is chosen such that $\langle
\lambda | \varphi_0 \rangle$ is real for all $| \lambda \rangle$
and satisfies $\langle \lambda | \varphi_0 \rangle \neq 0$. The
operator $\Phi_{2}$ is defined by $\Phi_{2} \equiv \sum_{\lambda}
c_{\lambda}|\lambda \rangle \langle \lambda |$ with $c_{\lambda}
\equiv N^{-{1 \over 2}}{\langle \lambda| \varphi_0 \rangle}^{-1}$.

It is noted that $\Phi_2$ is Hermitian and commutes with the
operator $R$. The square of $\Phi_2$ equals the inverse of $\Phi$,
i.e., $\Phi_2^2 = \Phi^{-1}$. The operator $\Phi_2$ becomes the
inverse square-root of $\Phi$ only when all $\langle \lambda |
\varphi_0 \rangle$ are real and positive.

{\it Proof of the Proposition}. We need to prove that the POM
elements defined in equations (\ref{eqn:POM}) and
(\ref{eqn:Gamma}) are indeed POM elements and satisfy the
necessary and sufficient conditions in equations
(\ref{eqn:opRelation}) and (\ref{eqn:positivity}). From equations
(\ref{eqn:POM}) and (\ref{eqn:Gamma}), we can prove that all
$\pi_k \geq 0$ as follows
\begin{equation}
\langle \phi | \pi_k | \phi \rangle = |\langle \phi | R^k \Phi_{2}
|\varphi_0 \rangle |^2 \geq 0, \hspace{1cm}\textrm{ for arbitrary
}k, |\phi\rangle. \label{eqn:piPositive}
\end{equation}
\noindent We can also {\it see} that $\pi_0 \geq 0$ by expanding
$\pi_0$ in the basis $\{|\lambda\rangle\}$, $\pi_0 = {1 \over N}
\sum_{\lambda,\lambda'}| \lambda \rangle\langle \lambda'|$. Under
the basis all the matrix elements of $\pi_0$ equals $1 / N$, thus
$\pi_0$ has only one non-vanishing eigenvalue $1$. The requirement
that all eigenvalues of $R$ are different guarantees that all the
POM elements sum into identity operator
\begin{eqnarray}
\sum_{k=0}^{N-1}\pi_k &=& \sum_{k}R^k
\pi_0 R^{-k} \nonumber \\
&=&{1 \over N}\sum_{k}\sum_{\lambda,\lambda'}({b_{\lambda} \over
b_{\lambda'}})^k|\lambda\rangle\langle\lambda'| \nonumber \\
&=&\sum_{\lambda}|\lambda\rangle\langle\lambda| \nonumber \\
&=&\hat{\mathbf{1}}. \label{eqn:sumAllPi}
\end{eqnarray}

We proceed to prove that the POM given by equation (\ref{eqn:POM})
does satisfy the necessary and sufficient conditions listed in
equations (\ref{eqn:opRelation}) and (\ref{eqn:positivity}). By
taking equation (\ref{eqn:POM}) into (\ref{eqn:opRelation}), we
find
\begin{eqnarray}
&\pi_k&(p_k\rho_k - p_j\rho_j)\pi_j \nonumber\\
&=&{1 \over
N}R^k\Phi_{2}|\varphi_0\rangle\langle\varphi_0|\Phi_{2}
(\rho_0R^{j-k}-R^{j-k}\rho_0)\Phi_{2}
|\varphi_0\rangle\langle\varphi_0|\Phi_{2} R^{-j}.
\label{eqn:proof1}
\end{eqnarray}
\noindent By using equations (\ref{eqn:Phiinlambda}) and
(\ref{eqn:Rinlambda}) and that all $\langle \lambda |\rho_0
|\lambda' \rangle$ and $\langle \lambda | \varphi_0 \rangle$ are
real, we derive the following identity thus prove that equation
(\ref{eqn:proof1}) actually equals zero
\begin{eqnarray}
&\langle&\varphi_0|\Phi_{2}(\rho_0 R^{j-k}-R^{j-k} \rho_0)
\Phi_{2} |\varphi_0\rangle \nonumber \\
&=&\sum_{\lambda, \lambda'} c_{\lambda} c_{\lambda'} \langle
\varphi_0|\lambda\rangle\langle\lambda|\rho_0|\lambda'\rangle\langle
\lambda'|\varphi_0\rangle(b_{\lambda'}^{j-k}-b_{\lambda}^{j-k})
\nonumber \\
&=&0. \label{eqn:proof2}
\end{eqnarray}

The condition in equation (\ref{eqn:positivity}) is proved as
follows. First we observe that $\sum_{k}\pi_k\rho_k$ is Hermitian
by
\begin{eqnarray}
\sum_{k}\pi_k\rho_k &=& \sum_{k}R^k \pi_0\rho_0R^{-k} \nonumber \\
&=&{1 \over N}\sum_k\sum_{\lambda,\lambda', \lambda''}
({b_{\lambda} \over b_{\lambda''}})^k|\lambda\rangle\langle
\lambda'|\rho_0|\lambda''\rangle\langle\lambda''| \nonumber \\
&=&\sum_{\lambda,\lambda'}|\lambda\rangle\langle\lambda'|\rho_0|
\lambda\rangle\langle\lambda| \nonumber \\
&=&\sum_{k}\rho_k\pi_k. \label{eqn:pirho}
\end{eqnarray}
\noindent Therefore the operators $(\sum_{k}\pi_k\rho_k-\rho_j)$
are also Hermitian for all $j$. By sandwiching
$(\sum_{k}\pi_k\rho_k-\rho_0)$ using an arbitrary state $|\phi
\rangle$, we have
\begin{eqnarray}
&&\langle\phi|\sum_{k}\pi_k\rho_k-\rho_0|\phi\rangle \nonumber
\\&=& \sum_{\lambda,\lambda'}(|\langle\phi|\lambda\rangle|^2-\langle
\phi|\lambda' \rangle \langle \lambda|\phi \rangle)\langle
\lambda' |\rho_0 | \lambda\rangle \nonumber \\
&=&{1\over 2} \sum_{\lambda,\lambda'} (|\langle \phi|\lambda
\rangle|^2 + |\langle\phi|\lambda'\rangle|^2 -\langle
\phi|\lambda' \rangle \langle \lambda|\phi \rangle - \langle
\phi|\lambda \rangle \langle \lambda' |\phi \rangle) \langle
\lambda' |\rho_0 | \lambda \rangle \nonumber \\
&=& {1\over 2} \sum_{\lambda,\lambda'} \varepsilon_{\lambda
\lambda'}(\phi) \langle \lambda'| \rho_0 |\lambda \rangle \geq 0,
\label{eqn:PositiveCond} \\
&\varepsilon_{\lambda \lambda'}&(\phi) \equiv (\langle \lambda
|\phi \rangle - \langle \lambda' | \phi \rangle)(\langle \lambda
|\phi \rangle - \langle \lambda' | \phi \rangle)^{\dagger}\geq 0.
\label{eqn:epsilon}
\end{eqnarray}
\noindent From equation (\ref{eqn:PositiveCond}) we conclude that
$(\sum_{k} \pi_k \rho_k- \rho_0)$ is a Hermitian operator and
nonnegative-definite. This then leads to the fact that $(\sum_{k}
\pi_k\rho_k- \rho_j)$ are also nonnegative-definite and Hermitian
for all possible $j$ since $( \sum_{k}\pi_k \rho_k- \rho_j )= R^j
(\sum_{k} \pi_k \rho_k- \rho_0) R^{\dagger j}$.

\section{Examples}
It is instructive to consider some examples of symmetric quantum
signals and solve for the optimal discrimination strategies by
using the proposition provided in the previous section.
\noindent{\it Ex. 1: Signals as pure quantum states}

Although our proposition aims at providing optimal discrimination
strategy for mixed quantum states, it can also be applied to the
case that has only symmetric pure quantum states. Given that
$\rho_k = R^{k}(\theta) |\Psi_0 \rangle \langle \Psi_0| R^{\dagger
k}(\theta)$ with
\begin{eqnarray}
|\Psi_0\rangle &=& \left(%
\begin{array}{c}
  1 \\
  0 \\
\end{array}%
\right), \\
R(\theta)&=& \left(%
\begin{array}{cc}
  \cos[{\theta \over 2}] & -\sin[{\theta \over 2}] \\
  \sin[{\theta \over 2}] & \cos[{\theta \over 2}] \\
\end{array}%
\right), \hspace{1cm}\theta = {2 \pi \over N}. \label{eqn:R2}
\end{eqnarray}
\noindent we find that $R(\theta)$ has two eigenstates $|
\lambda_1 \rangle ={1 \over \sqrt{2}}(1, -i)$, $|\lambda_2 \rangle
= {1 \over \sqrt{2}}(1, i)$ with eigenvalues $\lambda_1 = e^{i
\theta/2}$ and $\lambda_2 = e^{-i \theta/2}$, respectively. The
matrix elements $\langle \lambda | \rho_0 | \lambda' \rangle$ are
found to be real and nonnegative for all $\lambda$ and $\lambda'$
in the basis $\{|\lambda_1 \rangle, |\lambda_2 \rangle \}$. On the
other hand, we may choose the operator $\Gamma_0 \equiv
|\varphi_0\rangle \langle \varphi_0|$ by assigning $|\varphi_0
\rangle = |\Psi_0\rangle$ so that all $\langle \lambda | \varphi_0
\rangle$ are positive real numbers. With such choice of
$|\varphi_0\rangle$, the operator $\Phi_2$ becomes the inverse
square-root of $\Phi$ ($\Phi_2 = \Phi^{-{1 \over 2}})$. Therefore,
we have
\begin{eqnarray}
\Phi &=& \sum_{k=0}^{N-1}R^{k} |\Psi_0 \rangle \langle \Psi_0 |
R^{\dagger k} = {N \over 2} \hat{\textbf{1}}, \label{eqn:purePhi}\\
\pi_k &=& \Phi^{-{1 \over 2}}|\Psi_k \rangle \langle \Psi_k |
\Phi^{-{1 \over 2}} = {2 \over N}|\Psi_k \rangle \langle \Psi_k |,
\label{eqn:purepik} \\
(P_{error})_{min}&=& 1 - |\langle \Psi_0|\Phi^{- {1 \over
2}}|\Psi_0 \rangle |^2 = 1 - {2 \over N}. \label{eqn:purePerror}
\end{eqnarray}
\noindent Equations (\ref{eqn:purePhi}-\ref{eqn:purePerror}) are
exactly the same results obtained in the literatures
\cite{Qdetect,Symmetric}. In this example, our method is identical
to the square-root measurement \cite{Symmetric}.

\noindent{\it Ex. 2: Signals as mixed quantum states (I)}

Consider three symmetric mixed quantum states that satisfy
equations (\ref{eqn:rho}, \ref{eqn:RN1}) with
\begin{equation}
\rho_0 = \left(%
\begin{array}{cc}
  1/3 & 0 \\
  0 & 2/3 \\
\end{array}%
\right). \label{eqn:diagonalRho}
\end{equation}
\noindent The rotation operator $R(\theta)$ is also given in
equation (\ref{eqn:R2}) with $\theta = 2\pi/3$. We find that if we
choose $\{|\lambda_1 \rangle = {1 \over \sqrt{2}}(-i, 1),
|\lambda_2 \rangle ={1 \over \sqrt{2}}(i, 1) \}$ as the basis that
spans the Hilbert space of the signal states, all the matrix
elements of $\rho_0$ will be real and nonnegative in the basis.
According to the proposition in the previous section, the pure
quantum state $|\varphi_0 \rangle$ must be chosen such that all
$\langle \lambda|\varphi_0 \rangle$ are nonzero and real. It is
easy to see that any pure quantum state $(a |\lambda_1\rangle + b
|\lambda_2\rangle)$ with nonzero real coefficients $a, b$ that
satisfy $a^2+b^2=1$ could be a candidate for $|\varphi_0\rangle$.
By choosing $|\varphi_0\rangle = {1 \over \sqrt{2}}(
|\lambda_1\rangle +|\lambda_2\rangle)$, we have $\Phi={3 \over
2}\hat{\textbf{1}}$ and
\begin{eqnarray}
\pi_0 &=&
{2 \over 3}\left(%
\begin{array}{cc}
  0 & 0 \\
  0 & 1 \\
\end{array}%
\right), \label{eqn:Ex2Pi}\\
(P_{errer})_{min}&=& 1- \textrm{tr}(\pi_0 \rho_0) = {5 \over 9}.
\label{eqn:PerrEx2}
\end{eqnarray}

Although we have just obtained the results by using the
proposition, we can also solve for the optimal measurement
strategy in a {\it direct} way. Let us expand the mixed quantum
state $\rho_0$ in terms of Pauli matrices $\{\sigma_1, \sigma_2,
\sigma_3\}$ and the identity operator $\hat{\textbf{1}}_2$ in the
spin-$1/2$ Hilbert space, $\rho_0 = ({1 \over 2}\hat{\textbf{1}}_2
- {1 \over 3} \sigma_3)$. Consider that $R$ is the rotation about
the $\hat{2}$-direction by an angle $\theta = 2\pi/3$, therefore
the optimal POM elements $\pi_k$ should be of the following
general forms:
\begin{equation}
\pi_k = R^k (b_0 \hat {\textbf{1}}_2 +2b_1 \sigma_1+ 2b_3
\sigma_3) R^{\dagger k}, \hspace{1cm} k=0,1,2. \label{eqn:piEx2}
\end{equation}
\noindent where $b_0$, $b_1$ and $b_2$ are the coefficients to be
determined. It is found that $b_0 = 1/3$ by requiring
$\sum_{k}\pi_k = \hat{\textbf{1}}_2$. By considering that all the
POM elements are Hermitian and nonnegative we get the constraint
on $b_1$ and $b_3$, $\sqrt{b_1^2+b_3^2}\leq 1/3$. We then use the
constraint to find the minimum of the error probability
$P_{error}$
\begin{eqnarray}
P_{error} &=& 1 - {1 \over 3}\sum_{k=0}^2 \textrm{tr}(\pi_k
\rho_k) \nonumber \\
&=& 1 -\textrm{tr}(\pi_0 \rho_0) \nonumber \\
&=& {2 \over 3} + {b_3 \over 3} \geq {5 \over 9}.
\label{eqn:lowbound}
\end{eqnarray}
\noindent From equation (\ref{eqn:lowbound}), the probability of
error $P_{error}= 5/9$ is optimal at $b_1=0$ and $b_3=-1/3$, which
is exactly the same measurement strategy as described in equation
(\ref{eqn:Ex2Pi}).

\noindent{\it Ex. 3: Signals as mixed quantum states (II)}

In the previous examples we discussed the optimal discrimination
among single-qubit quantum states. Here we would like to discuss
how to discriminate the mixed quantum states with higher
dimensions of Hilbert space.

Let $|0\rangle$, $|1\rangle$ and $|2\rangle$ be the trine states
$(1, 0)$, $(1/2, \sqrt{3}/2)$ and $(1/2, -\sqrt{3}/2)$ that
respect the $Z_3$ symmetry $R(\theta=2\pi/3)$, respectively.
\begin{eqnarray}
R(\theta)^3 &=& -\hat{\textbf{1}}_2,
\hspace{2.0cm}\theta={2\pi\over 3}
\nonumber \\
|k\rangle &=& R(\theta)^k |0\rangle, \hspace{1cm}
k=0,1,2.\label{eqn:012}
\end{eqnarray}
\noindent The quantum states to be discriminated are not
single-qubit states but the two-qubit quantum states that also
respect the $Z_3$ symmetry
\begin{eqnarray}
\rho_0 &=& {1 \over 2}\{(|1\rangle\otimes|2\rangle)(\langle
1|\otimes \langle 2|) + (|2\rangle\otimes|1\rangle)(\langle
2|\otimes \langle 1|) \}, \nonumber \\
\rho_1 &=& {1 \over 2}\{(|2\rangle\otimes|0\rangle)(\langle
2|\otimes \langle 0|) + (|0\rangle\otimes|2\rangle)(\langle
0|\otimes \langle 2|) \}, \nonumber \\
\rho_2 &=& {1 \over 2}\{(|0\rangle\otimes|1\rangle)(\langle
0|\otimes \langle 1|) + (|1\rangle\otimes|0\rangle)(\langle
1|\otimes \langle 0|) \}. \label{eqn:2qubitRho}
\end{eqnarray}
\noindent It is obvious that these quantum states are reducible
mixed states. They can be decomposed into direct sums of spin-$1$
and spin-$0$ parts by ${1\over 2}\otimes{1 \over 2} = 1 \oplus 0$.
The spin-$1$ Hilbert subspace is spanned by the spin-$1$ states
$|1,1\rangle$, $|1,0\rangle$ and $|1,-1\rangle$. The spin-$0$
subspace is of dimension one, and is spanned by the spin-$0$ state
$|0,0\rangle$. By assigning $|1,1\rangle = (1,0,0)$, $|1,0\rangle
= (0,1,0)$ and $|1,-1\rangle = (0,0,1)$ we can rewrite the mixed
signals in the following matrix forms
\begin{eqnarray}
\rho_k &=& R_3^k \tilde{\rho}_0 R_3^{\dagger k} \oplus {3 \over 8}
|0,0\rangle \langle0,0|, \hspace{2cm} k=0,1,2.
\nonumber \\
\tilde{\rho_0}&\equiv& \left(%
\begin{array}{ccc}
  1/16 & 0 & -3/16 \\
  0 & 0 & 0 \\
  -3/16 & 0 & 9/16 \\
\end{array}%
\right) , \nonumber \\
R_3&\equiv& \left(%
\begin{array}{ccc}
  \cos^2[{\theta \over 2}] & {1 \over \sqrt{2}}\sin[\theta] &
  \sin^2[{\theta \over 2}] \\
  {-1 \over \sqrt{2}}\sin[\theta] & \cos[\theta] & {1 \over \sqrt{2}}
  \sin[\theta] \\
  \sin^2[{\theta \over 2}] & {-1 \over \sqrt{2}}\sin[\theta] &
  \cos^2[{\theta \over 2}] \\
\end{array}%
\right),  \hspace{1cm} \theta= {2\pi \over 3}.
\label{eqn:Ex3States}
\end{eqnarray}
\noindent It is easy to verify that the rotation operator $R_3$
does respect the $Z_3$ symmetry by $R_3^3 = \hat{\textbf{1}}$, and
has three different eigenvalues. The best measurement strategy
$\{\pi_k\}$ can also be decomposed into direct sums as $\pi_k =
\tilde{\pi}_k \oplus {1 \over 3}|0,0\rangle \langle0,0| $. Again,
the operators $\tilde{\pi}_k$ denote the POM elements in spin-$1$
subspace.

The probability of error can also be viewed as being contributed
from different Hilbert subspaces as
\begin{eqnarray}
P_{error} &=& 1-\sum_{subspaces} (\sum_{k} \textrm{tr}(p_k
\tilde{\pi}_k \tilde{\rho}_k)) \nonumber \\
&=& 1- {1 \over 3}\sum_{k}(\tilde{\pi}_k \tilde{\rho}_k)_{spin 1}
-{1 \over 8}, \label{eqn:Ex3Prror}
\end{eqnarray}
\noindent where $1/8$ in equation (\ref{eqn:Ex3Prror}) comes from
tracing over spin-$0$ subspace. As seen from equation
(\ref{eqn:Ex3Prror}), only the measurement in spin-$1$ subspace
needs to be optimized. We will solve the optimization problem by
using the proposition given in the previous section.

First we note that the rotation operator $R_3$ has three different
eigenvalues $1$, $e^{-i 2\pi/3}$ and $e^{i 2\pi/3}$ with
normalized eigenstates $|\lambda_1\rangle={1 \over
\sqrt{2}}(1,0,1)$, $|\lambda_2\rangle={1 \over 2}(-1,i\sqrt{2},1)$
and $|\lambda_3\rangle={1 \over 2}(-1,-i\sqrt{2},1)$,
respectively. In the basis formed by $\{|\lambda_1 \rangle,
|\lambda_2 \rangle, |\lambda_3 \rangle \}$, all matrix elements of
$\tilde{\rho}_0$ are real and nonnegative. According to the
proposition, the pure quantum state $|\varphi_0 \rangle$ must be
chosen such that $\langle \lambda | \varphi_0 \rangle$ are real
and nonzero for all possible $|\lambda \rangle$. A convenient
choice for $|\varphi_0 \rangle$ is $|\varphi_0 \rangle = (0, 0,
1)$. We then obtain the operator $\Phi_2$ in spin-$1$ Hilbert
subspace
\begin{equation}
\Phi_2 = {1 \over \sqrt{6}} \left(%
\begin{array}{ccc}
  1+\sqrt{2} & 0 & 1-\sqrt{2} \\
  0 & 2\sqrt{2} & 0 \\
  1-\sqrt{2} & 0 & 1+\sqrt{2}\\
\end{array}%
\right). \label{eqn:Ex3Phi}
\end{equation}
\noindent Therefore we get
\begin{eqnarray}
\tilde{\pi}_0 &=& \Phi_2 \Gamma_0 \Phi_2 \nonumber
\\ &=&
{1 \over 6}\left(%
\begin{array}{ccc}
  3-2\sqrt{2} & 0 & -1 \\
  0 & 0 & 0 \\
  -1 & 0 & 3+2\sqrt{2} \\
\end{array}%
\right),
\end{eqnarray}
\noindent and the optimal error probability is $(3-\sqrt{2})/6$.
This result coincides with our previous calculation that uses von
Neumann measurement for signal discrimination \cite{oldpaper}.
This coincidence is reasonable. Since the dimension of Hilbert
space is larger than the number of the signal states in this
example, both the optimal POM and the optimal von Neumann
measurement may have the same optimal probability of error.


\subsection*{Acknowledgments}

This work was supported in part by National Science Council of
Taiwan. 

\end{document}